\documentclass[12pt]{article}
\usepackage{graphicx}
\usepackage{jheppub}
\usepackage{ytableau}
\usepackage{pgf,pgfarrows,pgfnodes}
\usepackage{graphics}
\usepackage{hyperref}
\title{The light asymptotic limit of conformal
blocks in Toda field theory}
\author[a]{Hasmik Poghosyan,}
\author[a]{Rubik Poghossian}
\author[a,b]{and Gor Sarkissian}
\affiliation[a]{Yerevan Physics Institute\\
Alikhanian Br. 2, 0036 Yerevan, Armenia}
\affiliation[b]{Department of Physics, \ Yerevan State University,\\
Alex Manoogian 1, 0025\, Yerevan.
Armenia}
\emailAdd{hasmikpoghos@gmail.com}
\emailAdd{poghos@yerphi.am}
\emailAdd{gor.sarkissian@ysu.am}
\abstract{We compute the light asymptotic limit of $A_{n-1}$ Toda conformal blocks by using the AGT correspondence. We show that for  certain class of CFT blocks the corresponding Nekrasov partition functions in this limit are simplified drastically being represented as a sum of a restricted class of Young diagrams. In the particular case of $A_{2}$ Toda we also compute the corresponding conformal blocks using conventional CFT techniques finding a perfect agreement with the results obtained from the Nekrasov partition functions.  }

\begin{document}
\maketitle
\newcommand{\ie}{{\it i.e.\ }}
\def\bea{\begin{eqnarray}}
\def\eea{\end{eqnarray}}
\def\a{\alpha}
\def\b{\beta}
\def\g{\gamma}
\def\G{\Gamma}
\def\d{\delta}
\def\D{\Delta}
\def\e{\epsilon}
\def\z{\zeta}
\def\th{\theta}
\def\k{\kappa}
\def\l{\lambda}
\def\m{\mu}
\def\n{\nu}
\def\r{\rho}
\def\s{\sigma}
\def\t{\tau}
\def\f{\phi}
\newpage
\section{Introduction}

Semiclassical limits play important role since they link quantum physics
to the Lagrangian approach. In the Liouville and Toda field theories there are
three semiclassical limits: mini-superspace \cite{Braaten:1982yn,Braaten:1983np,Thorn:2002am,Fateev:2007ab}, the light and heavy \cite{Seiberg:1990eb,Zamolodchikov:1995aa,Fateev:2007ab}.
All three asymptotics are the large central charge  limits.
The difference comes in the treatment of the primary fields.
In the minisuperspace limit
one considers a limit where only the zero mode dynamics survives. In this limit the Liouville and Toda field theories
reduce to the corresponding quantum mechanical problems  \cite{Braaten:1982yn,Braaten:1983np,Fateev:2007ab}.
In the light asymptotic limit one keeps the conformal dimensions fixed. Then the correlation functions are given by the
finite dimensional path integral over solutions of the equations of motion with a vanishing energy-momentum tensor.
And finally in the heavy asymptotic limit the conformal dimensions blow up, scaling as the classical action
and correlation functions are given by the exponential of the action evaluated over the singular solutions.

To be more specific recall that primary fields in the  Liouville and Toda field theories
are related to the vertex operators $V_{\alpha}=e^{i\alpha \phi}$. The spectrum is given by $\alpha={Q\over 2}+iP$.
In the light asymptotic limit we set $\alpha=\eta_l b$ and keep $\eta_l$ fixed for $b\to 0$, whereas
in the heavy asymptotic limit we take $\alpha={\eta_h\over b}$ and hold $\eta_h$ fixed again for $b\to 0$.
In the minisuperpsace limit one should take for some of the vertex operators $\alpha=\eta_m b$ and for some
$P=\eta_m b$.

These semiclassical limits were used in \cite{Thorn:2002am,Zamolodchikov:1995aa,Fateev:2007ab} to relate
 the quantum three-point functions in the Liouville and Toda theories
with the corresponding classical  actions.
The heavy asymptotic limit plays an important role in the quantum uniformization program \cite{Ginsparg:1993is}.
In papers \cite{Fateev:2010za,Hadasz:2006vs,Menotti:2006gc,Menotti:2006tc}
these techniques were generalized to the boundary Liouville and Toda theories.
The heavy and light asymptotic limits were reconsidered  in \cite{Harlow:2011ny} also for complex solutions of the analytically continued Liouville theory.
In \cite{Poghosyan:2015oua} both limits were used in the Liouville field theory with defects.
Both limits have recently proved to be very useful also to test AdS/CFT correspondence \cite{Hijano:2015qja,Alkalaev:2015wia,Fitzpatrick:2015zha}.

Discovery of AGT correspondence \cite{Alday:2009aq,Wyllard:2009hg,
Alba:2010qc,Fateev:2011hq} relating 2d CFT conformal blocks to
the Nekrasov partition function \cite{Lossev:1997bz,Nekrasov:2002qd} in ${\cal N}=2$ supersymmetric
gauge theory  provides powerful tools to investigate
CFT correlators using gauge theory methods or alternativly to apply
advanced CFT methods in gauge theory (see e.g. \cite{Alday:2009fs,
Poghossian:2009mk}. The essential point here is the fact that there are
explicit combinatorial formulas for the Nekrasov partition function
\cite{Flume:2002az,Bruzzo:2002xf}, which now can be successfully applied in 2d CFT.

The heavy asymptotic limit in the AGT context was considered in \cite{Marshakov:2010fx,Piatek:2013ifa,Poghossian:2016rzb}.
 In \cite{Mironov:2009qn,Fateev:2011qa,Hama:2013ama} the light asymptotic limit was used to test the AGT correspondence.

 Investigation of the light asymptotic limit of the Nekrasov functions for $U(2)$ $N=2$ gauge theory in \cite{Mironov:2009qn,Hama:2013ama}
revealed a very interesting fact, that in this limit only the Young 
diagrams consisting of a single row contribute.
This could be anticipated since in the light asymptotic limit the infinite Virasoro symmetry reduces
to SL(2) algebra whose representations are classified with one row Young tableaux.

The sum over these one-row Young tableaux yields the Gauss Hypergeometric function, which is the well known light asymptotic limit
of the Virasoro conformal blocks  \cite{Zamolodchikov:1985ie}.

In \cite{Fateev:2011qa} the AGT correspondence was tested in the light asymptotic limit between conformal blocks of $A_2$  Toda field theory and Nekrasov functions of  $U(3)$ $N=2$ gauge theory.
Namely some conformal blocks for  $A_2$  Toda field theory were computed  in this limit and the result was checked
 against the corresponding limit of the Nekrasov functions up to the 
5-th order terms. But such an elegant picture as it was found
for the Liouville field theory was missing.

In this paper we consider the light asymptotic limit of the $U(n)$ Nekrasov partition functions for an arbitrary $n$.
We find that for the certain choice of fields the Nekrasov partition functions in the light asymptotic limit are
simplified drastically and given by the sum over Young diagrams having at most $n-1$ rows.
We compute the corresponding $W_3$ conformal block using the light asymptotic integral representation and found
perfect agreement with the two-row Nekrasov partition functions.
Note that in the light asymptotic limit the $W_n$ symmetry reduces to $SL(n)$ group \cite{Bowcock:1991zk,Fateev:2011qa} and this already hints
on the existence of the limiting procedure where survive only Young diagrams corresponding to the
$SL(n)$ representations.

The paper is organized as follows.
In section \ref{tlal} we compute the light asymptotic limit of the Nekrasov partition functions.
In subsection \ref{npfof} we review the necessary facts on the Nekrasov partition functions.
In subsection \ref{patcft} we review Toda conformal field theory and the AGT relation.
In subsection \ref{lalita} we explain the details on the light asymptotic limit and show
that choosing the data as it is specified in eq. (\ref{eta23}) and (\ref{au_light}) truncates the Nekrasov functions
in the light asymptotic limit to the sum over Young tableaux containing at most $n-1$ rows.
In subsection \ref{npfsymlat} we compute the Nekrasov partition function in the light asymptotic limit.
The formula (\ref{centralrez}) is our main result.
In section \ref{fpbw3} we compute the corresponding conformal block in $A_2$ Toda field theory using
that in the light asymptotic limit conformal blocks admit an integral representations.
In appendix A the details of the integral calculation in section \ref{fpbw3} are delivered.

\section{The light asymptotic limit of the Nekrasov partition functions}
\label{tlal}

\subsection{The Nekrasov partition functions of ${\cal N}=2 $ SYM theory}
\label{npfof}
Consider ${\cal N}=2 $ SYM theory with gauge group $U(n)$ and $2 n$
fundamental (more precisely $n$ fundamental plus $n$ anti-fundamental)
hypermultiplets in $\Omega $-background. The instanton part of the
partition of this theory can be represented as
\begin{eqnarray}\label{nekfu}
Z_{inst}=\sum_{\vec{Y}}F_{\vec{Y}}q^{|\vec{Y}|},
\end{eqnarray}
where $\vec{Y}$ is an array of $n$ Young diagrams,
 $|\vec{Y}|$ is the total
number of  boxes and $q$ is the instanton counting parameter related to
the gauge coupling in a standard manner. The coefficients $F_{\vec{Y}}$
are given by
\begin{eqnarray}
F_{\vec{Y}}=\prod_{u=1}^n\prod_{v=1}^n
\frac{Z_{bf}(a_u^{(0)},\varnothing\mid a_v^{(1)},Y_v)
Z_{bf}(a_u^{(1)},Y_u\mid a_v^{(2)},\varnothing)}
{Z_{bf}(a_u^{(1)},Y_u\mid a_v^{(1)},Y_v)}\,,
\label{F}
\end{eqnarray}
where
\begin{eqnarray}
\label{Zbf}
Z_{bf}(a,\lambda\mid b,\mu)=&\\
\displaystyle\prod_{s\in\lambda}\big(a-b-\epsilon_1L_{\mu}(s)+
\epsilon_2(1+A_{\lambda}(s))\big)&
\displaystyle\prod_{s\in\mu}\big(a-b+\epsilon_1(1+L_{\lambda}(s))
-\epsilon_2A_{\mu}(s)\big)\,.\nonumber
\end{eqnarray}
\begin{figure}
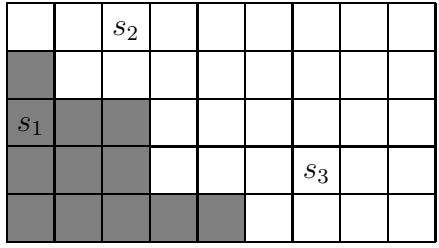

  \centering
\ytableausetup{nosmalltableaux}
\centering
\begin{ytableau}
*(white) &  &s_2 & & & & & &\\
*(gray) &  & & & & & & &\\
*(gray)s_1 & *(gray) &*(gray) & & & & & &\\
*(gray) & *(gray) &*(gray) & & & & s_3& &\\
*(gray) & *(gray) & *(gray)&*(gray) &*(gray)& & & &
\end{ytableau}
\caption{Arm and leg length with respect to a Young diagram (pictured in gray): $A(s_1)=1$, $L(s_1)=2$, $A(s_2)=-2$, $L(s_2)=-3$, $A(s_3)=-2$, $L(s_3)=-4$. }
\label{YD1}
\end{figure}
Here $A_{\lambda}(s)$ and $L_{\lambda}(s)$ are correspondingly the arm-length and leg-length
of the square $s$ towards the Young tableau $\lambda$, defined as oriented vertical and horizontal distances
of the square $s$ to outer boundary of the Young tableau $\lambda$ (see Fig.\ref{YD1}).

Let us clarify our conventions on gauge theory parameters $a_u^{(0,1,2)}$,
$u=1,2,\ldots,n $. The parameters $a_u^{(1)}$ are expectation values
of the scalar field in vector multiplet. Without loss of generality
we'll assume that the ``center of mass" of these expectation values is zero
\bea\label{au1}
{\bar a}^{(1)}=\frac{1}{n}\sum_{u=1}^n a_u^{(1)}=0\,.
\eea
In fact this is not a loss of generality since a nonzero center of mass can be absorbed by shifting hypermultiplet
masses. Furthermore $a_u^{(0)}$ ($a_u^{(2)}$) are the masses of fundamental
(anti-fundamental) hypers. Finally the $\epsilon_1$, $\epsilon_2$ are the
$\Omega$-background parameters. Sometimes we will use the notation
$\epsilon=\epsilon_1+\epsilon_2$.

Due to AGT duality,
this partition function is directly related to specific
four point conformal block in 2d $A_{n-1}$ Toda field theory. Before
describing this relation let us briefly recall few facts about Toda theory.
\subsection{Preliminaries on $A_{n-1}$ Toda CFT and AGT relation}
\label{patcft}
These are 2d CFT theories which besides the spin $2$ holomorphic
energy momentum $W^{(2)}(z)\equiv T(z)$ are endowed with additional
higher spin
$s=3, 4,\ldots,n $ currents $W^{(3)}$, \ldots , $W^{(n)}$ with Virasoro
central charge conventionally parameterised as
\[
c=n-1+12 Q^2\, ,
\]
where the vector ``background charge"
\[
Q=\rho (b+1/b)
\]
with $\rho $ being the Weyl vector of the algebra $A_{n-1}$ and $b$ is the
dimensionless coupling constant of Toda theory. In what follows it would
be convenient to represent the roots, weights and Cartan elements of $A_{n-1}$
as $n$-component vectors with the usual Kronecker scalar product, subject
to the condition that sum of components is zero. Of course this is equivalent
to more conventional representation of these quantities as diagonal traceless
$n\times n$ matrices with the pairing given by trace. In this representation
the Weyl vector is given by
\bea\label{weylvec}
\rho =\left(\frac{n-1}{2},\frac{n-3}{2},\ldots ,\frac{1-n}{2}\right)\quad {\rm or}\quad \rho_u={n+1\over 2}-u
\eea
and for the central charge we`ll get
\[
c=(n-1)(1+n(n+1)q^2)\, ,
\]
where for the later use we have introduced the parameter
\[
q=b+\frac{1}{b} \, .
\]
For further reference let us quote here explicit expressions for the highest weight $\omega_1$ of the first fundamental representation and for its complete set of weights $h_1, \ldots,h_n$ ($h_1=\omega_1)$
\bea
&&(\omega_1)_k=\delta_{1,k}-1/n\nonumber \, ;\\
&&(h_l)_k=\delta_{l,k}-1/n \, .
\eea
The primary fields $V_\alpha$ (here we concentrate only on left moving
holomorphic parts) are parameterized by vectors $\alpha $ with
vanishing center of mass. Their conformal weights are given by
\bea
h_{\alpha}=\frac{\alpha (2Q-\alpha)}{2} \, .
\label{dim_gen}
\eea
In what follows a special role is played by the fields $V_{\lambda \omega_1}$
with the dimensions:
\bea
h_{\lambda \omega_1}=\frac{\lambda(n-1)}{2} \left(q-\frac{\lambda}{n}\right) \, .
\label{dim_spec}
\eea
A four point block:
\bea\label{block gen}
\langle V_{\alpha^{(4)}}(\infty)V_{\lambda^{(3)} \omega_1}(1)
V_{\lambda^{(2)}\omega_1}(q)
V_{\alpha^{(1)}}(0)\rangle_{\alpha}=
q^{h_{\alpha}-h_{\alpha^{(1)}}-h_{\alpha^{(2)}\omega_1}}
{\cal F}_{\alpha}\left[\begin{array}{cc}
\lambda^{(3)} \omega_1&\lambda^{(2)}\omega_1\\
\alpha^{(4)}&\alpha^{(1)} \end{array}\right](q)\,,\qquad
\eea
where  $\alpha$
specifies the W-family running in s-channel, is closely related to the
gauge partition function $Z_{inst}$  see (\ref{zincon}) (AGT relation). First of all,
the instanton counting parameter $q$ gets identified with the cross
ratio of insertion points in CFT block as it was already
anticipated in (\ref{block gen}) and the Toda parameter $b$ is related to
$\Omega$-background parameters via
\bea
b=\sqrt{\frac{\epsilon_1}{\epsilon_2}}\,.
\label{bepsilon}
\eea
\begin{figure}[t]
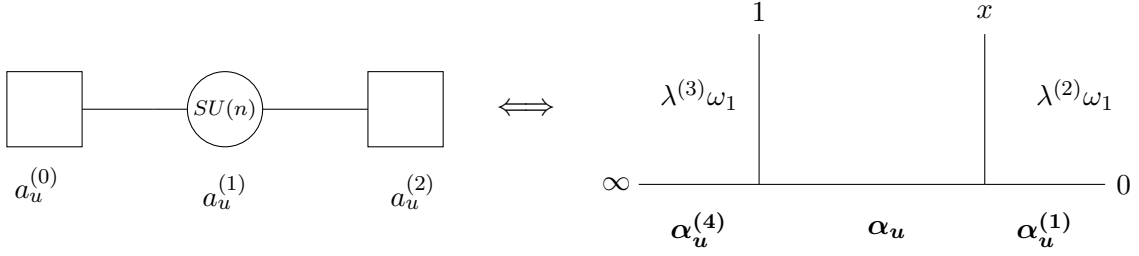

\begin{pgfpicture}{0cm}{0cm}{15cm}{4cm}

\pgfcircle[stroke]{\pgfpoint{3cm}{2.8cm}}{0.5cm}
\pgfputat{\pgfxy(3,2.8)}{\pgfbox[center,center]{\scriptsize{$SU(n)$}}}
\pgfputat{\pgfxy(3,1.7)}{\pgfbox[center,center]{\small{$a_u^{(1)}$}}}
{\color{black}\pgfrect[stroke]{\pgfpoint{0.1cm}{2.3 cm}}{\pgfpoint{1cm}{1cm}}}
\pgfputat{\pgfxy(0.5,1.8)}{\pgfbox[center,center]{\small{$a_u^{(0)}$}}}
{\color{black}\pgfrect[stroke]{\pgfpoint{4.9cm}{2.3cm}}{\pgfpoint{1cm}{1cm}}}
\pgfputat{\pgfxy(5.5,1.7)}{\pgfbox[center,center]{\small{$a_u^{(2)}$}}}
\pgfline{\pgfxy(1.1,2.8)}{\pgfxy(2.05,2.8)}
\pgfline{\pgfxy(2.05,2.8)}{\pgfxy(2.5,2.8)}
\pgfline{\pgfxy(3.5,2.8)}{\pgfxy(4.9,2.8)}
\pgfputat{\pgfxy(7,2.8)}{\pgfbox[center,center]{$\Longleftrightarrow$}}
\pgfline{\pgfxy(8.5,1.8)}{\pgfxy(10.1,1.8)}
\pgfline{\pgfxy(10.1,1.8)}{\pgfxy(10.1,3.8)}
\pgfline{\pgfxy(10.1,1.8)}{\pgfxy(13.1,1.8)}
\pgfline{\pgfxy(13.1,1.8)}{\pgfxy(13.1,3.8)}
\pgfline{\pgfxy(13.1,1.8)}{\pgfxy(14.7,1.8)}
\pgfputat{\pgfxy(11.8,1.2)}{\pgfbox[center,center]{\small{$\boldsymbol{\alpha_u} $}}}
\pgfputat{\pgfxy(9.3,3)}{\pgfbox[center,center]{\small{$\lambda^{(3)}\omega_1$}}}
\pgfputat{\pgfxy(9.3,1.2)}{\pgfbox[center,center]{\small{$\boldsymbol{\alpha_u^{(4)}} $}}}
\pgfputat{\pgfxy(14.3,3)}{\pgfbox[center,center]{\small{$\lambda^{(2)}\omega_1$}}}
\pgfputat{\pgfxy(13.9,1.2)}{\pgfbox[center,center]{\small{$\boldsymbol{\alpha_u^{(1)}} $}}}
\pgfputat{\pgfxy(8.2,1.8)}{\pgfbox[center,center]{\small{$\infty$}}}
\pgfputat{\pgfxy(10.1,4.1)}{\pgfbox[center,center]{\small{$1$}}}
\pgfputat{\pgfxy(13.1,4.1)}{\pgfbox[center,center]{\small{$x$}}}
\pgfputat{\pgfxy(14.95,1.8)}{\pgfbox[center,center]{\small{$0$}}}
\pgfclearendarrow
\end{pgfpicture}
\caption{On the left: the quiver diagram for the conformal $U(n)$ gauge theory. On the right: the diagram of the conformal block for the dual Toda field theory.}
\label{figAGT}
\end{figure}
The map between the gauge parameters in (\ref{nekfu}) and conformal block parameters in (\ref{block gen})
should be established from the following rules (see Fig.\ref{figAGT}). To formulate them we define the rescaled gauge parameters
\bea\label{renga}
A_u^{(0)}=\frac{a_u^{(0)}}{\sqrt{\epsilon_1\epsilon_2}}\,;\quad\quad A_u^{(1)}=\frac{a_u^{(1)}}{\sqrt{\epsilon_1\epsilon_2}}\,;\quad
\quad A_u^{(2)}=\frac{a_u^{(2)}}{\sqrt{\epsilon_1\epsilon_2}}\,.
\eea
\begin{itemize}
\item
The differences between the ``centers of mases" of the successive rescaled gauge parameters (\ref{renga}) give the charges of  the ``vertical" entries of the conformal block:
\bea
\bar{A}^{(1)}-\bar{A}^{(0)}=\frac{\lambda^{(3)}}{n}\, ;\quad\quad \bar{A}^{(2)}-\bar{A}^{(1)}=\frac{\lambda^{(2)}}{n}\,.
\eea
\item
The rescaled gauge parameters with the subtracted centers of masses give the momenta of the ``horizontal" entries of the conformal block:
\bea\label{hormom}
&&A_u^{(0)}-\bar{A}^{(0)}=Q_u-\alpha_u^{(4)}\, ;\\ \nonumber
 &&A_u^{(1)}-\bar{A}^{(1)}=Q_u-\alpha_u\,; \\ \nonumber
&&A_u^{(2)}-\bar{A}^{(2)}=Q_u-\alpha_u^{(1)}\, .
\eea
\end{itemize}
Using (\ref{au1}), (\ref{weylvec}) and (\ref{renga})-(\ref{hormom}) we obtain the relation
between the gauge and conformal parameters:
\begin{eqnarray}
\frac{a_u^{(0)}}{\sqrt{\epsilon_1\epsilon_2}}&=&-\alpha_u^{(4)}
-\frac{\lambda^{(3)}}{n}+q\left(\frac{n+1}{2}-u\right);\nonumber\\
\frac{a_u^{(1)}}{\sqrt{\epsilon_1\epsilon_2}}&=&-\alpha_u+
q\left(\frac{n+1}{2}-u\right);\nonumber\\
\frac{a_u^{(2)}}{\sqrt{\epsilon_1\epsilon_2}}&=&-\alpha_u^{(1)}+
\frac{\lambda^{(2)}}{n}+q\left(\frac{n+1}{2}-u\right).
\label{au}
\end{eqnarray}
With all these preparations one can write the AGT correspondence between the Nekrasov function defined in (\ref{nekfu})
and the conformal block in (\ref{block gen}) (see \cite{Wyllard:2009hg,
Fateev:2011hq}):
\bea\label{zincon}
Z_{inst}=(1-q)^{\lambda^{(3)}\left(q-{\lambda^{(2)}\over n}\right)}{\cal F}_{\alpha}\left[\begin{array}{cc}
\lambda^{(3)} \omega_1&\lambda^{(2)}\omega_1\\
\alpha^{(4)}&\alpha^{(1)} \end{array}\right](q)\,.
\eea
\subsection{Light asymptotic limit}
\label{lalita}
In this paper we are interested in so called "light" asymptotic
limit i.e. the central charge is sent to infinity (i.e. $b\rightarrow 0$)
while keeping the dimensions finite. It follows from (\ref{dim_gen}) that
to reach this limit one can simply put
\bea
\alpha_u^{(1)}=b \eta_u^{(1)};\qquad\qquad \alpha_u^{(4)}=b \eta_u^{(4)}
; \qquad\qquad \alpha_u=b \eta_u
\label{eta14}
\eea
keeping all the parameters $\eta$ finite.
As for the parameters $\lambda$ of the special fields $V_{\lambda\omega_1}$,
there are two inequivalent alternatives:\\
(i) $\lambda=b \eta $\\
or\\
(ii) $n q-\lambda=b \eta $.\\
Though in both cases the conformal dimension takes the same value
(see eq. (\ref{dim_spec}))
\[
h=\frac{\eta (n-1)}{2}\,\,,
\]
these fields are not identical, which can be seen e.g. from the fact
that the zero mode eigenvalues of odd W-currents for these fields have
the same absolute values but opposite signs (in fact the fields
$V_{\lambda\omega_1}$ and $V_{(nq-\lambda)\omega_1}$ can be considered
as conjugate to each other in the usual sense, since their two point
function is non-zero).

In this paper we will investigate in great detail the case when
$V_{\lambda_3\omega_1}$ is a light field of type (i) while
$V_{\lambda_2\omega_1}$ is of type (ii). In other words we set
\bea
\lambda^{(3)}=b \eta^{(3)}; \qquad\qquad n q-\lambda^{(2)}=b \eta^{(2)}\, .
\label{eta23}
\eea
For such choice we will see below, that the corresponding instanton sum
simplifies drastically and leads to a simple explicit expression for the
conformal block. Note that this choice is very convenient since the prefactor in front of
conformal block in (\ref{zincon}) now goes to $1$ in the light asymptotic limit.
The opposite case when two special fields
are of the same type, has been investigated in \cite{Fateev:2011qa} in particular case of
$A_2$ Toda. In the case considered in \cite{Fateev:2011qa} the above mentioned prefactor survives.

\begin{figure}[]
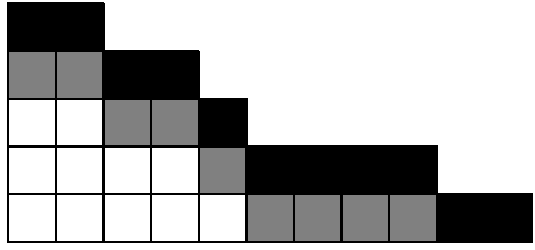

  \centering
\ytableausetup{centertableaux}
\ytableaushort
{\none }
* {2,4,5,9,11}
* [*(black)]{2,2+2,4+1,5+
4,9+2}
* [*(gray)]{2,2,2+2,4+1,5+4}
\caption{This picture shows that there are $Y_{v,1}$ boxes such that $A_{Y_v}=0$
(painted black) and $Y_{v,2}$ boxes with $A_{Y_v}=1$ (painted grey).}
\label{YD2}
\end{figure}

Coming back to our case of interest using (\ref{eta23}),
(\ref{eta14}) we can rewrite the AGT map (\ref{au}) as
\begin{eqnarray}
\label{au_light}
a_u^{(0)}&=&-\epsilon_1\left(\eta_u^{(4)}+\frac{\eta^{(3)}}{n}\right)+
\epsilon\left(\frac{n+1}{2}-u\right) \, ;\nonumber\\
a_u^{(1)}&=&-\epsilon_1\eta_u+
\epsilon\left(\frac{n+1}{2}-u\right)\nonumber \, ;\\
a_u^{(2)}&=&-\epsilon_1\left(\eta_u^{(1)}+
\frac{\eta^{(2)}}{n}\right)
+\epsilon\left(\frac{n+3}{2}-u\right) \, .
\end{eqnarray}
In view of (\ref{bepsilon}) the small $b$ limit is equivalent
to $\epsilon_1\rightarrow 0$. Hence we are interested in the
$\epsilon_1 \rightarrow 0$ limit of (\ref{F}).
  We will see that the degree of $\epsilon_1$ (denote it by $N$)
  is non-negative for
 arbitrary array of Young diagrams $Y_v$ and that the degree $N=0$
 (hence a finite non-zero limit exists) if and only if each Young
 diagram $Y_v$ ($v=1,2,\ldots, n$) has at most $v-1$ rows. \\
  From (\ref{F}) we see that
\begin{eqnarray}
N=n_1+n_2-n_3
\label{n}
\end{eqnarray}
with $n_1$, $n_2$ being the $\epsilon_1$ degrees of the first and second
factors in the numerator of (\ref{F}) respectively and $n_3$ is the
$\epsilon_1$ degree of its denominator.\\
Let us derive $n_1$, using (\ref{Zbf}) for
$Z_{bf}(a_u^{(0)},\varnothing \mid a_v^{(1)},Y_v)$
and inserting (\ref{au_light})
we'll get
\begin{eqnarray}
\label{Zbf0n}
&Z_{bf}(a_u^{(0)},\varnothing\mid a_v^{(1)},Y_v)=&\\ \nonumber
&\prod_{s\in Y_v}\big(\epsilon_1(1+L_{\varnothing}(s)+
\eta_v-\eta_u^{(4)}-\frac{\eta^{(3)}}{n})
+\epsilon(v-u)-\epsilon_2A_{Y_v}(s)\big).&
\end{eqnarray}
A factor in (\ref{Zbf0n}) contributes to the degree of  $\epsilon_1$ if
its part proportional to $\epsilon_2$ vanishes. Evidently this happens
when $A_{Y_v}(s)=v-u$. Since the box $s\in Y_v $, $A_{Y_v}(s)\ge 0$,
we see that when $v=1$ the only admissible value for $u$ is $u=1$.
It is obvious from Fig. \ref{YD2} that there are exactly $Y_{1,1}$ boxes in $Y_1$ for
which the arm-length vanishes (here and below we denote by $Y_{v,i}$
the number of boxes in the $i$'th row of diagram $Y_v$). When $v=2$,
there are two admissible values $u=1$ or $u=2$. As in the previous case
the number of the boxes with zero arm-length (case $u=2$) is equal $Y_{2,1}$.
Similarly, a simple inspection shows that the number of boxes with
unit arm-lengths (case $u=2$) are equal to $Y_{2,2}$. This analysis
can be easily continued for other values of $v$ with result summarized
in the table below
 \begin{center}

\begin{tabular}{ l |c|c|c|c c c |c| r }
              & u=1 & u=2 & u=3 &  & \ldots  & & u=n \\
\hline
v=1           & $Y_{1,1}$ & &  &  &  &  &   \\
\hline
v=2           & $Y_{2,2}$ & $Y_{2,1}$ &  &  &   &  &    \\
\hline
v=3           & $Y_{3,3}$ & $Y_{3,2}$ & $Y_{3,1}$ &  &   &  &
\\
\hline
\ldots          & &  &  &  &  \ldots &  &    \\
\hline
v=n           & $Y_{n,n}$ & $Y_{n,n-1}$ &$Y_{n,n-2}$ &  &  \ldots
&  &  $Y_{n,1}$  \\
\hline
\end{tabular}
\end{center}
Obviously the degree $n_1$ is nothing but the  sum of all entries of
this table.
\begin{eqnarray}
n_1=\displaystyle\sum_{u=1}^{n}\displaystyle\sum_{k=1}^uY_{u,k}\,.
\label{n1}
\end{eqnarray}
With almost identical arguments it is possible to show that $n_2=n_1$.
Finally, an analogous consideration for the degree $n_3$ gives
\begin{eqnarray}
n_3=\displaystyle\sum_{u=1}^{n}\displaystyle\sum_{k=1}^uY_{u,k}+
\displaystyle\sum_{u=1}^{n}\displaystyle\sum_{k=1}^{u-1}Y_{u,k}\,.
\label{n2}
\end{eqnarray}
Thus for the total degree (\ref{n}) we get
\begin{eqnarray}
N=\displaystyle\sum_{u=1}^{n}Y_{u,u}\,.
\end{eqnarray}
Each term here is non-negative and in order to get a vanishing
total degree $n=0$, the array of Young diagrams should satisfy
 the conditions $Y_{1,1}=Y_{2,2}=\cdots Y_{n,n}=0$, which
means that each Young diagram $Y_u$  consists of at most $u-1$ rows.

\subsection{Nekrasov partition function of ${\cal N}=2 $ SYM theory
in the light asymptotic limit}
\label{npfsymlat}
Now our purpose is to derive $F_{\vec{Y}}$  explicitly in the
light asymptotic limit.
To do this let us study the first factor in the
numerator of (\ref{F}) which, according to (\ref{Zbf}) and
(\ref{au_light}), is given by
\begin{eqnarray}
\label{aZbf0n}
&Z_{bf}(a_u^{(0)},\varnothing\mid a_v^{(1)},Y_v)=&\\ \nonumber
&\prod_{s\in Y_v}\big(\epsilon_1(1+L_{\varnothing}(s)+
\eta_v-\eta_u^{(4)}-\frac{\eta^{(3)}}{n})
+\epsilon(v-u)-\epsilon_2A_{Y_v}(s)\big).&
\end{eqnarray}
Let $Y_v^{(1)}$ be the set of such boxes $s$ of the Young diagram
$Y_v$ (with at most $v-1$ rows) that  the coefficient
of $\epsilon_2$ vanishes in the respective factor
of (\ref{aZbf0n})), i.e.
\begin{eqnarray}
v-u-A_{Y_v}(s)=0\,.
\label{v-u-A=0_condition}
\end{eqnarray}
This can happen only when $i \equiv v-u\ge 0$.
Thus for the part of (\ref{aZbf0n}) under discussion we get
\begin{eqnarray}
&Z_{bf}(a_u^{(0)},\varnothing\mid a_v^{(1)},Y_v^{(1)})=
\prod_{s\in Y_v^{(1)}}\epsilon_1(1+L_{\varnothing}(s)+
\eta_v-\eta_u^{(4)}-\frac{\eta^{(3)}}{n}+v-u)\,.&
\label{Zbf_part1_initial}
\end{eqnarray}
We have already seen in previous chapter that there
are exactly $Y_{v,i+1}$ boxes, satisfying
(\ref{v-u-A=0_condition}). These boxes are distributed in
$Y_v$ in such a way that there is a single box on
$j$-th column (denote it by $s_j$) for each $j=1,
\ldots,Y_{v,i+1}$ (see Fig.\ref{YD2}).
Taking into account that $L_{\varnothing}(s_j)=-j$,
we can rewrite (\ref{Zbf_part1_initial}) as
\begin{eqnarray}
\label{Zbf11}
&Z_{bf}(a_u^{(0)},\varnothing\mid a_v^{(1)},Y_v^{(1)})=
\prod_{j=1}^{Y_{v,i+1}}
\epsilon_1(\eta_v-\eta_{v-i}^{(4)}-\frac{\eta^{(3)}}{n}+1-j+i).&
\end{eqnarray}
Now let's look on the alternative case of
the set $Y_v^{(2)}$ of those boxes which do not satisfy
(\ref{v-u-A=0_condition}) so that in the related
factors we can safely set $\epsilon_1=0$. Again from
(\ref{Zbf0n}) we'll get
\begin{eqnarray}
 &Z_{bf}(a_u^{(0)},\varnothing\mid a_v^{(1)},Y_v^{(2)})=
\prod_{s\in Y_v}\epsilon_2\big(v-u-A_{Y_v}(s)\big).&
\end{eqnarray}
Carefully examining the cases $v-u-A_{Y_v}(s)>0$ and
$v-u-A_{Y_v}(s)<0$ separately we get
\begin{eqnarray}
\label{Zbf12}
&\prod_{u=1}^n\prod_{v=1}^n
Z_{bf}(a_u^{(0)},\varnothing\mid a_v^{(1)},Y_v^{(2)})=& \\
&\prod_{v=2}^n \prod_{i=1}^{v-1}
\big((-)^{n-v-1-i}(n-v-1-i)!(v-i)!
\epsilon_2^{(n-1)}\big)^{Y_{v,i}}.&\nonumber
\end{eqnarray}
Combining (\ref{Zbf11}) with (\ref{Zbf12})
we obtain
\begin{eqnarray}
\label{F1}
&\prod_{u=1}^n\prod_{v=1}^n
Z_{bf}(a_u^{(0)},\varnothing\mid a_v^{(1)},Y_v)=& \\ \nonumber
&\prod_{v=2}^{n}\prod_{i=0}^{v-2}\prod_{j=1}^{Y_{v,i+1}}
\epsilon_1\big(\eta_v-\eta_{v-i}^{(4)}
-\frac{\eta^{(3)}}{n}+1-j+i\big)\, \times&\\ \nonumber
&\prod_{v=2}^n \prod_{i=1}^{v-1}
\big((-
)^{n-v-1-i}(n-v-1-i)!(v-i)!\epsilon_2^{(n-1)}\big)^{Y_{v,i}}.&
\end{eqnarray}
Similar arguments for the second factor in the denominator
of (\ref{F}) lead to the expression
\begin{eqnarray}
\label{F2}
&\prod_{u=1}^n\prod_{v=1}^n
Z_{bf}(a_u^{(1)},Y_u\mid a_v^{(0)},\varnothing)=& \\
&\prod_{u=2}^n\prod_{i=1}^{u-1}
\big(\epsilon_2^{n-1}(-)^i
i!(n-1-i)!\big)^{Y_{u,u-i}}\, \times&\nonumber \\
&\prod_{u=2}^{n}\prod_{i=0}^{u-2}
\prod_{j=1}^{Y_{u
,i+1}}\epsilon_1\big(\eta_{u-i}^{(1)}
-\eta_u+\frac{\eta^{(2)}}{n}+j-i-1\big).&\nonumber
\end{eqnarray}
The derivation of the denominator of (\ref{F}) though
somewhat lengthier but still is quite straightforward
and leads to
\begin{eqnarray}
\label{F3}
&\prod_{u=1}^n\prod_{v=1}^n
Z_{bf}(a_u^{(1)}Y_u\mid a_v^{(1)}Y_v)=&\\ \nonumber
&\prod_{l=1}^{n-1}\prod_{v=1}^{n-l}(-\epsilon_1)^{Y_{v+l,l}}
\prod_{k=l}^{v+l-1}\prod_{i=1+Y_{v+l,k+1}}^{Y_{v+l,k}}
(\eta_{v+l}-\eta_{v}+l+Y_{v,k-l+1}-i)\, \times&\\\nonumber
&\prod_{l=0}^{n-2}\prod_{v=l+2}^n\epsilon_1^{Y_{v,l+1}}
\prod_{k=l+1}^{v-1}\prod_{i=1+Y_{v,k+1}}^{Y_{v,k}}
(\eta_v-\eta_{v-l}+l+1+Y_{v-l,k-l}-i)\, \times&\\ \nonumber
&\prod_{v=2}^n \prod_{i=1}^{v-1}
(\epsilon_2^{n-1}(-)^{i-1}(i-1)!(n-i)!)^{Y_{v,v-i}}
((-)^{n-v-1-i}(n-v-1-i)!(v-i)!\epsilon_2^{(n-1)})^{Y_{v,i}},&
\nonumber
\end{eqnarray}
where the products on the second (third) line comes from the
terms $u<v$ ($u>v$) and the last line results in from diagonal
$u=v$ terms.
Notice that, as we have already proved earlier, the order
in $\epsilon_1$ of the numerator and the denominator coincide
safely providing a finite $\epsilon_1 \rightarrow 0$ limit.
Also dependence of the ratio in $\epsilon_2$ disappears (as it should
from scaling arguments).
Inserting  (\ref{F1}), (\ref{F2}) and (\ref{F3}) in (\ref{F})
for $F_{\vec{Y}}$ in the light asymptotic limit we finally get
\begin{eqnarray}
\label{centralrez}
&F_{\vec{Y}}=
\prod_{u=2}^n \prod_{v=2}^{u}
\left(\frac{u-v+1}{n-u+v-1}\right)^{Y_{u,v-1}}&\\
&\frac{
\prod_{i=0}^{Y_{u,u-v+1}-1}\left(-\eta_u+\eta_{v}^{(4)}+
\frac{\eta^{(3)}}{n}-u+v+i\right)
\left(\eta_u-\eta_v^{(1)}-\frac{\eta^{(2)}}{n}+u-v-i\right)
}
{\prod_{k=u-v+1}^{u-1}\prod_{i=Y_{u,k+1}}^{Y_{u,k}-1}
(\eta_u-\eta_{v-1}+u-v+Y_{v-1,k+v-u}-i)(\eta_u-\eta_v+u-v+Y_{v,k+v-u}-i)}\,\,\,.&
\nonumber
\end{eqnarray}
Let us consider the particular cases when $n=2$ (Liouville)
and $n=3$ separately.

When $n=2$ we have a single sum
\bea
 &{\cal F}_{Liouv}=\sum_{n=0}^{\infty}\frac{\left(\eta_2^{(4)}
 +\eta_1+\frac{\eta^{(3)}}{2}\right)_n\left(\eta_2^{(1)}
 +\eta_1+\frac{\eta^{(2)}}{2}\right)_n}{n! \left(2\eta_1\right)_n}\,\,x^n&\nonumber\\
&=\,_2F_1\left(\eta_2^{(1)}+\eta_1+
\frac{\eta^{(2)}}{2}\, ,\eta_2^{(4)}+\eta_1
+\frac{\eta^{(3)}}{2}\, , 2\eta_1;\,x\right)\,,&
\eea
where $\, _2F_1(a,b;c;x)$ is the Gauss hyper-geometric function.
This is a well known result in Liouville theory \cite{Zamolodchikov:1985ie,Mironov:2009qn,Hama:2013ama,Fateev:2011qa}.

When $n=3$ we get
  \bea
  \label{W3_block}
&{\cal F}_{W_3}=\sum_{i,j,l=0}^{\infty}(-)^l 2^{j-i}
\,x^{2l+i+j}&\\
&\times \left(\frac{\eta^{(3)}}{3}-\eta_2+\eta_2^{(4)}\right)_i
 \left(\frac{\eta^{(3)}}{3}-\eta_3+\eta_2^{(4)}-1\right)_l
 \left(\frac{\eta^{(3)}}{3}-\eta_3+\eta_3^{(4)}\right)_{j+l}  &\nonumber\\
 &\times\frac{ \left(\frac{\eta^{(2)}}{3}-\eta_2+\eta_2^{(1)}\right)_i
 \left(\frac{\eta^{(2)}}{3}-\eta_3+\eta_2^{(1)}-1\right)_l
 \left(\frac{\eta^{(2)}}{3}-\eta_3+\eta_3^{(1)}\right)_{j+l}
 }{i!j!l! \left(\eta_1-\eta_2\right)_i \left(\eta_1-\eta_3-1\right)_l
\left(\eta_2-\eta_3\right)_l \left(\eta_2-\eta_3-i-1\right)_l
\left(\eta_2-\eta_3+l-i\right)_j }\,\,,&\nonumber
 \eea
This formula completes the result of \cite{Fateev:2011qa}
where the light four-point function of $W_3$-theory has been
computed in the case when both the second and the third insertions were
 light primaries of the same sort:
 \bea
\lambda^{(3)}=b \eta^{(3)}; \qquad\qquad \lambda^{(2)}=b \eta^{(2)}\, ,
\label{eta223}
\eea
whereas (\ref{W3_block})
 is obtained with the choice specified in (\ref{eta23}).
In the next section we present an alternative calculation
of (\ref{W3_block}) based on the integral representation of the conformal blocks in the light asymptotic limit used in \cite{Fateev:2011qa}.

\section{Light asymptotic limit for the four point block in $W_3$
}
\label{fpbw3}
It has been shown in \cite{Fateev:2011qa} that the multi-point conformal
blocks of the $W_3$ theory in the light asymptotic limit can be
constructed in the terms of $sl(3)$ three-point invariant
functions. For the details we refer the reader to the original
paper. Here we'll introduce the necessary notations and state
the relevant results.

It is well known that the $sl(3)$ generators can be represented as operators acting on the
triple of the isospin variables $\vec{Z}=(w,x,y)$. To construct
a multi-point block one should multiply several three-point
functions then identify pairs of isospin variables corresponding
to the internal states and integrate them out with
an appropriate measure. At the end one specializes the
external leg variables putting
\bea
\vec{Z}=\left(\frac{1}{2}\,z^2,z,z\right)\,,
\label{z_imbedding}
\eea
where $z$ is the insertion point.

In particular the four-point block can be represented as
\bea
\mathcal{F}=\int_C \,d^3\vec(Z)_s
\mathcal{E}_1(j_2,j_1,J_s^\omega|Z_2,Z_1,Z_s)
\mathcal{E}_2(j_3,j_4^\omega,J_s^{*\omega}|Z_3,Z_4,Z_s)\,,
\eea
where $\mathcal{E}_1$ and $\mathcal{E}_2$ are the
appropriate three point invariants given by
\footnote{We have different three point invariants,
since the second and third light fields
are of different kinds as specified in (\ref{eta23}). The case of
fields of the same kind is analysed in \cite{Fateev:2011qa}.}
\bea
&&\mathcal{E}_1(j_1,j_2,j_3|Z_1,Z_2,Z_3)=
\chi_{123}^{-J}\r_{12}^{-J-r_2+s_3}
\r_{13}^{-J-r_3+s_2}\r_{23}^{J-s_2}
\r_{32}^{J-s_3}\,;\nonumber\\
&&\mathcal{E}_2(j_1,j_2,j_3|Z_1,Z_2,Z_3)
=\s_{123}^J\r_{21}^{J+r_3-s_2}
\r_{31}^{J+r_2-s_3}\r_{23}^{-J-r_3}\r_{32}^{-J-r_2}
\eea
with
\bea
&&\rho _{ij}=y_i \left(x_i-x_j\right)-\left(w_i-w_j\right)\,;
\nonumber\\
&&\sigma _{ijk}=x_i w_j-w_i x_j-x_i w_k+w_i x_k-w_j x_k+x_j w_k\,;\nonumber\\
&&\chi _{ijk}=y_i w_j-w_i y_j+y_i y_j
\left(x_i-x_j\right)-y_i w_k+w_i y_k+y_i y_k
\left(x_k-x_i\right)\nonumber\\
&&\qquad-w_j y_k+y_j w_k+y_j y_k \left(x_j-x_k\right),
\eea
the quantities $j=(r,s)$, $j^*=(2-r,2-s)$, $j^\omega=(s,r)$
(see \cite{Fateev:2011qa}) specify the primary fields
and are related to the charge vectors $\eta_u$ introduced
in section (\ref{lalita}) as
\bea
r=\eta_1-\eta_2\,; \qquad s=\eta_2-\eta_3\,;\qquad
\eta_1+\eta_2+\eta_3=0
\label{rs}
\eea
and, finally,
\bea
J=(h_2,j_1+j_2+j_3)=\frac{1}{3}\,\,(s_1+s_2+s_3-r_1-r_2-r_3)\,.
\eea
Due to (\ref{rs}) and (\ref{eta14}), (\ref{eta23}) for our case we have
\bea
&r_s=\eta_1-\eta_2\, ; \qquad
s_s=\eta_2-\eta_3\, ; &\nonumber\\
&r_1=\eta^{(1)}_1-\eta^{(1)}_2\, ; \qquad
s_1=\eta^{(1)}_2-\eta^{(1)}_3\, ;& \nonumber\\
&r_4=\eta^{(4)}_1-\eta^{(4)}_2\, ; \qquad
s_4=\eta^{(4)}_2-\eta^{(4)}_3\, ; &\nonumber\\
&s_2=\eta^{(2)}\,; \qquad
r_2=0\, ;&\nonumber\\
&r_3=\eta^{(3)}\,;
\qquad s_3=0\,.&
\eea
As usual, using projective invariance we can specify
the insertion points as\\
$(z_4,z_3,z_2,z_1)\rightarrow (\infty,1,x,0)$, see Fig.\ref{figAGT}.
Under this specification, after dropping
out an unimportant constant (infinite ) factor,
$\mathcal{E}_2$ gets simplified
\bea
\mathcal{E}_2(j_3,j_4^\omega,J_s^{*\omega}|Z_3,Z_4,Z_s)=
\left (1-x_s\right)^{\frac{1}{3}(s_4+s_s-r_s-r_3-r_4)}
\rho_{s,3}^{\frac{s_4+s_s-r_s-r_3-r_4}{3}+r_4+r_s-2}\,.
\eea
Putting
\bea
&x_2\to z\, ;\qquad y_2\to z\, ;\qquad w_2\to \frac{z^2}{2}\, ;\qquad x_1\to 0\, ; &\nonumber \\
&y_1\to 0\, ;\qquad w_1\to 0\, ;\qquad x_3\to 1\, ;\qquad y_3\to 1\, ;\qquad w_3\to \frac{1}{2}\, , &\nonumber
\eea
as instructed in (\ref{z_imbedding})
and dropping out the usual factor \\$z^{h_{\alpha_s}-h_{\alpha^{(1)}}-h_{\lambda^{(2)}\omega_1}}
=z^{r_s+s_s-(r_1+s_1)-s_2}$, up to an unimportant constant multiplier
 we get the integral
\bea
\mathcal{F}=\int \,dx_s \,dy_s \,dw_s
\left(w_s-y_s \left(x_s-\frac{z}{2}\right)\right){}^{\frac{1}{3}
(r_1+s_s-s_1-s_2-r_s)}
 \qquad\qquad \qquad\\ \nonumber
\times w_s^{\frac{1}{3} (-r_1-s_s-2 s_1+s_2+r_s)}
\left(1-x_s\right){}^{\frac{1}{3} (-r_3-r_4+s_s+s_4-r_s)}
\left(w_s-x_s y_s\right){}^{\frac{1}{3} (-r_1-s_s+s_1+s_2-2
r_s)}\qquad \\ \nonumber
\times \left(w_s-z \left(x_s-\frac{z}{2}\right)\right){}^{\frac{1}{3}
(r_1-2 s_s+2 s_1-s_2-r_s)}
 \left(w_s-\left(x_s-1\right)
 y_s-\frac{1}{2}\right){}^{\frac{1}{3} (-r_3+2
 r_4+s_s+s_4+2r_s)-2}. \\ \nonumber
\eea
After the change of the variables
 \bea
 x_s\to \frac{x}{2 w}\, ;\qquad w_s\to \frac{1}{2 w}\, ;\qquad y_s\to \frac{y}{x y-w}
 \eea
we'll get
\bea
\label{int_final}
\mathcal{F}&=&\int_{\mathcal{C}} \,dx\,dy\,dw
w^{\frac{1}{3} (r_3+r_4+2 s_s-s_4+r_s)-2}
 (1-y z)^{\frac{1}{3} (r_1+s_s-s_1-s_2-r_s)} \\\nonumber
&\times& \left(w-\frac{x}{2}\right)^{\frac{1}{3} (-r_3-r_4+s_s+s_4-r_s)}
 \left(w z^2-x z+1\right)^{\frac{1}{3} (r_1-2 s_s+2 s_1-s_2-r_s)}
 \\\nonumber
 &\times&(x y-w)^{\frac{1}{3} (r_3-2 r_4-s_s-s_4+r_s)}
 (-w+(x-2) y+1)^{\frac{1}{3} (-r_3+2 r_4+s_s+s_4+2 r_s)-2}.
\eea
Here is the result of the integration (for the details of the
calculation see appendix:\ref{A})
\bea
 \label{w3block_FRtype}
&\mathcal{F}=\sum_{m,n,k=0}^\infty  \sum _{l=0}^m (-)^{k+l} 2^{n-m} z^{2 k+m+n}  &\nonumber \\
&\times \left(\frac{1}{3} \left(s_s+2 r_s-r_3-r_4+s_4\right)\right)_l \left(\frac{1}{3} \left(-s_s+r_s-r_1+s_1+s_2\right)\right)_m&\\
&\times \frac{  \left(\frac{1}{3} \left(2 s_s+r_s-r_1-2 s_1+s_2\right)\right)_{k+n} \left(\frac{1}{3} \left(2 s_s+r_s+r_3-2 r_4-s_4\right)\right)_{k+n} \left(\frac{1}{3} \left(2 s_s+r_s+r_3+r_4-s_4-3\right)\right)_{k-l+m}}{k! l! n! (m-l)! (r_s)_l (s_s)_{k-l+n} (s_s+r_s-1)_{k+m}}\,\,.& \nonumber
\eea
Though this expression looks different from (\ref{W3_block}) extensive
automatic calculation ensures that in fact they coincide.
\section{Discussion}
The methods developed in this paper can be extended to other cousins of the Liouville and Toda field theories,
like super Liouville and super Toda.
Also it is an important problem to compute the conformal blocks in the light asymptotic limit for an arbitrary
$n$ via the integral representation, as it was done for $A_2$ Toda field theory in the last section.

Let us also recall that in the light asymptotic limit the $W_n$ symmetry reduces to $SL(n)$ group.
This means that the function (\ref{centralrez}) in fact should have a group theoretical meaning.
It would be interesting to discover it.

\section*{Acknowledgments}
The work of R.P. and G.S. was partially supported by the Armenian SCS  grant 15T-1C308 and by ICTP OEA-AC-100 project.
The work of G.S. was also supported by
ICTP Network NET68.
The work of H.P. and G.S. was partially supported by the ANSEF grant hepth-4208.
\\
\vspace*{3pt}

\appendix
\section{The integral calculation}
\label{A}
For a suitable integration contour it is allowed to have boundaries ending
on branch points of the integrand of (\ref{int_final}). Another necessary
condition is that the result of integration should be analytic in insertion
point $z$. Here is an appropriate choice for contour $ \mathcal{C}$,
satisfying both requirements
\bea
\mathcal{C}:\quad y\in \left(\frac{w}{x},\frac{w-1}{x-2}\right)
\quad \text{then} \quad w\in \left(0,\frac{x}{2}\right)
\quad \text{then} \quad x\in (0,2)
\eea

There are to factors in integrand of (\ref{int_final})
which depend on $z$. Expanding the product of these
factors in powers of $z$ we get
\bea
\left( 1-yz\right)^{g}(1-xz+wz^2)^h=
\sum_{m,k,n=0}^{\infty}C_{m,k,n}z^{2k+m+n}x^nw^ky^m,
\eea
where
\bea
C_{m,k,n}=
(-)^{m+n}
\frac{\Gamma{(g+1)} \Gamma{(h+1)}}{m!k! n! \Gamma{(g-m+1)} \Gamma{(h-k-n+1)}}\,\,.
\eea
Inserting this into (\ref{int_final}) we'll find
\bea
&&\mathcal{F}=
\sum_{m,k,n=0}^{\infty}C_{m,k,n}z^{2k+m+n}
\int \,dx\,dy\,dw \\ \nonumber
&&y^m x^n w^{e+k}
\left(w-\frac{x}{2}\right)^f (x y-w)^A (-w+(x-2) y+1)^B,
\eea
where we introduced the notations
\bea
&&e=\frac{1}{3} (r_3+r_4+2 s_s-s_4+r_s)-2\,; \quad f= \frac{1}{3} (-r_3-r_4+s_s+s_4-r_s)\,;\nonumber\\
&&A= \frac{1}{3} (r_3-2 r_4-s_s-s_4+r_s)\,;\quad B=\frac{1}{3} (-r_3+2 r_4+s_s+s_4+2 r_s)-2\,;\nonumber\\
&&g=\frac{1}{3} (r_1+s_s-s_1-s_2-r_s)\,;\quad h=\frac{1}{3} (r_1-2 s_s+2 s_1-s_2-r_s)\,.
\label{ABefgh_notation}
\eea
Inserting binomial expansion
\bea
y^m=\sum_{l=0}^{m}\left( \frac{w}{x}\right)^{m-l}
\left( y-\frac{w}{x}\right)^{l} \binom{m}{l}
\eea
and shifting the variable $y\to y+\frac{w}{x} $ we'll get
the integral
\bea
&&\mathcal{F}=
\sum_{m,k,n=0}^{\infty}C_{m,k,n}z^{2k+m+n}
\int \,dx\,dy\,dw \\ \nonumber
&&\sum_{l=0}^{m}
x^A (x-2)^B x^n y^{A+l} w^{e+k} \left(w-\frac{x}{2}\right)^f \left(y-\frac{2 w-x}{x (x-2)}\right)^B \left(\frac{w}{x}\right)^{m-l}
\binom{m}{l}.
\eea
The result of integration over $y\in [0,\frac{2w-x}{x(x-2)}]$ is
 \bea
\mathcal{F}=
\sum_{m,k,n=0}^{\infty}C_{m,k,n}z^{2k+m+n}
\sum_{l=0}^{m}
2^l \binom{m}{l}
\frac{\Gamma{(B+1)} \Gamma{(A+l+1)}}{\Gamma{(A+B+l+2)}}
 \qquad \\ \nonumber
 \int \,dx\,dw \,
 (x-2)^{-A-l-1} x^{-B-m+n-1}
\left(w-\frac{x}{2}\right)^{A+B+f+l+1}
w^{e+k-l+m}.
\eea
Next we'll integrate over $w\in [0,x/2]$ and get
 \bea
\mathcal{F}=
\sum_{m,k,n=0}^{\infty}C_{m,k,n} 2^{n-m} z^{2k+m+n}
\sum_{l=0}^{m}
\binom{m}{l}
\frac{\Gamma{(B+1)} \Gamma{(A+l+1)}}{\Gamma{(A+B+l+2)}}
 \qquad \qquad \qquad\\ \nonumber
 \frac{\Gamma{(A+B+f+l+2)} \Gamma{(e+k-l+m+1)}}
 {\Gamma{(A+B+e+f+k+m+3)}}
 \int \,dx \,
 \left(1-\frac{x}{2}\right)^{-A-l-1} \left(\frac{x}{2}\right)^{A+e+f+k+n+1}.
\eea
The last integral over $x\in[0,2]$ is again of
Euler type so that the final result is
\bea
\mathcal{F}=
\sum_{m,k,n=0}^{\infty}C_{m,k,n} 2^{n-m} z^{2k+m+n}
\sum_{l=0}^{m}
\binom{m}{l}
\frac{\Gamma{(B+1)} \Gamma{(A+l+1)}}{\Gamma{(A+B+l+2)}}
 \qquad \qquad \qquad\\ \nonumber
 \frac{\Gamma{(A+B+f+l+2)} \Gamma{(e+k-l+m+1)}}
 {\Gamma{(A+B+e+f+k+m+3)}}
\frac{\Gamma{(-A-l)} \Gamma{(A+e+f+k+n+2)}}{\Gamma{(e+f+k-l+n+2)}}\,.
\eea
It remains to use (\ref{ABefgh_notation}) to arrive at (\ref{w3block_FRtype}).

\bibliographystyle{JHEP}
\providecommand{\href}[2]{#2}
\begingroup\raggedright

\endgroup

\end{document}